\documentclass[conference]{IEEEtran}
\usepackage{amssymb}
\usepackage[cmex10]{amsmath}
\usepackage{stfloats}
\usepackage{graphicx}
\usepackage{subfigure}
\usepackage{tabularx}
\usepackage{epsfig,epsf,color,balance,cite}
\usepackage{verbatim}
\usepackage{url}
\usepackage{bm}
\usepackage{mathrsfs}
\usepackage{amsthm}

\newtheorem{theorem}{Theorem}

\newtheorem{lemma}{Lemma}
\newtheorem{remark}{Remark}
\usepackage{algorithm}
\usepackage{algpseudocode}
\usepackage{graphicx}
\usepackage{epstopdf}
\begin{document}

\title{An Achievable Region for the Multiple Access Wiretap Channels with Confidential and Open Messages}

\author{
\IEEEauthorblockN{Hao Xu\IEEEauthorrefmark{1},
	              Giuseppe Caire\IEEEauthorrefmark{1},
and
	              Cunhua Pan\IEEEauthorrefmark{2},
}
\IEEEauthorblockA{\IEEEauthorrefmark{1}Faculty of Electrical Engineering and Computer Science, Technical University of Berlin, 10587 Berlin, Germany}
\IEEEauthorblockA{\IEEEauthorrefmark{2}School of Electronic Engineering and Computer Science, Queen Mary University of London, London E1 4NS, U.K.}
\IEEEauthorblockA{E-mail: xuhao@mail.tu-berlin.de; caire@tu-berlin.de; c.pan@qmul.ac.uk}

}

\maketitle

\begin{abstract}
This paper investigates the capacity region of a discrete memoryless (DM) multiple access wiretap (MAC-WT) channel where,
besides confidential messages,  the users have also open messages to transmit.
All these messages are intended for the legitimate receiver but only the confidential messages need to be protected from the 
eavesdropper. By using random coding, we find an achievable secrecy rate region, within which perfect secrecy 
can be realized, i.e., all users can communicate with the legitimate receiver with arbitrarily small probability of error, 
while the confidential information leaked to the eavesdropper tends to zero.
\end{abstract}

\IEEEpeerreviewmaketitle

\section{Introduction}
\label{section1}

Different from the key-based cryptographic techniques, information theoretic secrecy exploits the random propagation properties of radio channels to prevent eavesdroppers from extracting confidential information of authorized users, and has triggered considerable research interest recently \cite{yang2015safeguarding}.
The study of information theoretic secrecy in communications starts from several seminal papers \cite{wyner1975wire, leung1978gaussian, csiszar1978broadcast}.
In \cite{wyner1975wire}, Wyner considered a discrete memoryless (DM) channel with an eavesdropper which observes a stochastically degraded 
version of the output of the main channel, and aimed to maximize the transmission rate  to the legitimate 
receiver while keeping the eavesdropper as ignorant of the secret message as possible.
The trade-off between transmission rate and the equivocation of the eavesdropper was investigated, and the existence of a secrecy 
capacity was proven in \cite{wyner1975wire}. Based on \cite{wyner1975wire}, in \cite{leung1978gaussian} 
the achievable rate-equivocation region of a degraded Gaussian wiretap channel was investigated. 
In \cite{csiszar1978broadcast}, Wyner's work was extended to a non-degraded broadcast wiretap channel, and to a scenario including a confidential message for the legitimate receiver only and a common message intended for both the legitimate receiver and the eavesdropper.

Following the work in \cite{wyner1975wire, leung1978gaussian, csiszar1978broadcast}, the information theoretic secrecy problems for several other
channel models have been studied, including the multiple access wiretap (MAC-WT) channel \cite{tekin2008gaussian, ekrem2008secrecy, tekin2008general, nafea2019generalizing}. Both \cite{tekin2008gaussian} and \cite{ekrem2008secrecy} considered a MAC-WT channel with a weaker eavesdropper 
which sees a degraded version of the main channel.  
Reference \cite{tekin2008gaussian} developed an outer bound for the secrecy capacity region of the DM MAC-WT channel.
In \cite{ekrem2008secrecy}, two separate secrecy measures were first defined for a Gaussian MAC-WT channel, and achievable rate regions 
were provided for different secrecy constraints by using Gaussian inputs and stochastic encoders.
In \cite{tekin2008general} and \cite{nafea2019generalizing}, non-degraded MAC-WT channels were considered.
Specifically, in \cite{tekin2008general}, the authors extended the work of \cite{tekin2008gaussian} to a general Gaussian MAC-WT channel, and besides secret message, each user also had an open message to transmit.
An achievable  rate region for both secret and open rates was then provided. In \cite{nafea2019generalizing}, the MAC-WT channel with a DM main 
channel and different wiretapping scenarios were considered.

In this paper, we study the information theoretic secrecy problem for a general MAC-WT channel.
Different from \cite{tekin2008general}, which assumes Gaussian inputs and Gaussian channels, we consider DM channels.
Each user is assumed to have a secret message and an open message to transmit.
This constitutes a generalization of the results in \cite{nafea2019generalizing}, where each user only transmits a secret message.
By using random coding, we find an achievable secrecy rate region, where users can communicate with legitimate receiver with the arbitrarily small probability of error, while the confidential information leaked to the eavesdropper tends to zero.

Furthermore, we also show that the analogous achievable region given in \cite{tekin2008general}~\footnote{Although this 
derived for the Gaussian MAC with Gaussian inputs, it can be easily stated in terms of mutual information expressions and directly compared with our result.}
does not hold in general. In this sense, our result amends \cite[Theorem 1]{tekin2008general} which appears to be not correct in general. 

\section{Channel Model}
\label{section2}

\begin{figure}
	\centering
	\includegraphics[scale=0.40]{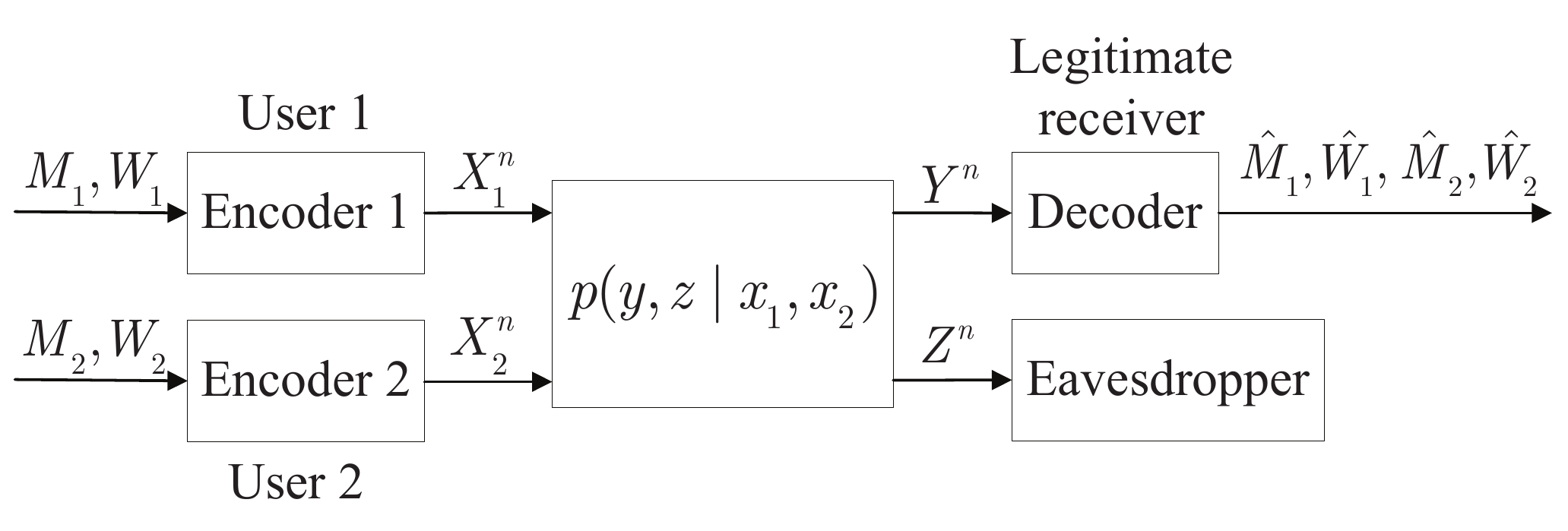}
	\caption{DM MAC-WT channel with an eavesdropper.}
	\label{Fig1}
\end{figure}

Consider a MAC wiretap communication system with two users, a legitimate receiver and an eavesdropper as shown in Fig. \ref{Fig1}.
Let ${\cal K} = \{1, 2\}$ denote the set of users.
Each user $k \in {\cal K}$ needs to send a secret message $M_k$ and an open message $W_k$ to the legitimate receiver.
We assume a DM wiretap channel $({\cal X}_1, {\cal X}_2, p(y,z|x_1, x_2), {\cal Y}, {\cal Z})$
where $x_1 \in {\cal X}_1$ and $x_2 \in {\cal X}_2$ are respectively channel inputs from user $1$ and user $2$, and $y \in {\cal Y}$ and $z \in {\cal Z}$ are respectively channel outputs at the legitimate receiver and the eavesdropper. 

Let $R_k^{\text s}$ and $R_k^{\text o}$ denote the rate of user $k$'s secret and open messages, respectively. Then, a $\left( 2^{n R_1^{\text s}}, 2^{n R_1^{\text o}}, 2^{n R_2^{\text s}}, 2^{n R_2^{\text o}}, n \right)$ secrecy code for the considered MAC-WT channel consists of
\begin{itemize}
	\item Four message sets: ${\cal M}_k = [1:2^{n R_k^{\text s}}]$ and ${\cal W}_k = [1:2^{n R_k^{\text o}}], \forall k \in {\cal K}$. 
	Messages $M_k$ and $W_k$ are uniformly distributed over ${\cal M}_k$ and ${\cal W}_k$, respectively. 
	\item Two randomized encoders: the encoder of user $k$ maps message pair $(M_k, W_k) \in {\cal M}_k \times {\cal W}_k$ to a codeword $X_k^n \in {\cal X}_k^n$.
	\item A decoder at the legitimate receiver which maps a received sequence $Y^n \in {\cal Y}^n$ to message pairs $\left( {\hat M}_k, {\hat W}_k \right) \in  {\cal M}_k \times {\cal W}_k, \forall k \in {\cal K}$.
\end{itemize}

The secrecy level of the MAC system is evaluated by the information leakage rate, which is defined as
\begin{equation}\label{leakage_rate}
R_{{\text E}, {\cal S}} = \frac{1}{n} I (M_{\cal S}; Z^n), ~\forall~ {\cal S} \subseteq {\cal K},
\end{equation}
where $M_{\cal S} = \left\{ M_k, \forall k \in {\cal S} \right\}$.
For perfect secrecy of all transmitted secret messages, we would like $R_{{\text E}, {\cal S}} \rightarrow 0, \forall {\cal S} \subseteq {\cal K}$.
Note that since messages $M_1$ and $M_2$ are independent, we have
\begin{align}\label{M1M2}
I (M_1, M_2; Z^n) & = I (M_1; Z^n) + I (M_2; Z^n | M_1)\nonumber\\
& \geq I (M_1; Z^n) + I (M_2; Z^n),
\end{align}
which indicates that if 
the leakage rate for all confidential messages vanishes, then the system is secure also for all possible message subsets.

The average probability of error is defined as
	\begin{align}\label{Pe}
	P_{\text e} = & P \left\{ \left( {\hat M}_1, {\hat W}_1, {\hat M}_2, {\hat W}_2 \right) \neq \left( M_1, W_1, M_2, W_2 \right) \right\}.
	\end{align}
	A rate tuple $(R_1^{\text s}, R_1^{\text o}, R_2^{\text s}, R_2^{\text o})$ is said to be achievable if for any $\delta > 0$ there exists a sequence of 
	$\left( 2^{n R_1^{\text s}}, 2^{n R_1^{\text o}}, 2^{n R_2^{\text s}}, 2^{n R_2^{\text o}}, n \right)$ codes for increasing $n$ such that
	\begin{align}
	& \lim_{n \rightarrow \infty} P_{\text e} \leq \delta,\\
	& \lim_{n \rightarrow \infty} R_{{\text E}, {\cal K}} \leq \delta.\label{RE}
	\end{align}

\section{Main Results}
\label{section3}

We use the short-hand notation $p(x_1)$ and $p(x_2)$ to indicate $P_{X_1}(x)$ with $x \in {\cal X}_1$ 
and $P_{X_2}(x)$ with $x \in {\cal X}_2$, respectively. Analogous short-hand notations are clear from the context. 
In this section, we state our main results.

\begin{theorem}\label{theorem1}
	Let $(X_1, X_2, Y, Z) \!\sim~\!p(x_1) p(x_2) p(y,z| x_1, x_2)$. Then, any rate tuple $\left(R_1^{\text s}, R_1^{\text o}, R_2^{\text s}, R_2^{\text o}\right)$ satisfying
	\begin{equation}\label{rate_region}
	\!\left\{\!\!\!
	\begin{array}{ll}
	\sum_{k \in {\cal S}} (R_k^{\text s} + R_k^{\text o}) \leq I(X_{\cal S}; Y| X_{\bar {\cal S}}), \forall \cal S \subseteq \cal K, \\
	\sum_{k \in {\cal S}} R_k^{\text s} \leq \left[I(X_{\cal S}; Y| X_{\bar {\cal S}}) - I(X_{\cal S}; Z) \right]^+\!\!, \forall \cal S \subseteq \cal K,\\
	R_1^{\text s} + R_2^{\text s} + R_k^{\text o} \leq \left[I(X_{\cal K}; Y) - I(X_{\bar k}; Z) \right]^+\!\!, \forall k \in \cal K,
	\end{array} \right.\!\!
	\end{equation}
	is achievable, where $X_{\cal S} = \left\{ X_k :  k \in {\cal S} \right\}$, $\bar {\cal S}$ is the complement 
	set of $\cal S$, i.e., $\bar {\cal S} = \cal K \setminus \cal S$, ${\bar k} = 1$ if $k=2$, and ${\bar k} = 2$ if $k=1$.
	Let ${\mathscr R} (X_1, X_2)$ denote the set of rate tuples satisfying (\ref{rate_region}).
	Then, the convex hull of the union of ${\mathscr R} (X_1, X_2)$ over all $p(x_1)  p(x_2)$ is an achievable secrecy rate region of the considered MAC wiretap channel.
\end{theorem}

\itshape \textbf{Proof:}  \upshape
	See Section \ref{section4}.
\hfill $\Box$

The result in Theorem \ref{theorem1} can be directly extended to the more general case with $K \geq 1$ users.
\begin{lemma}\label{lemma1}
	Denote ${\cal K} \!=\! \{1, \cdots, K \}$ and let $(X_{\cal K}, Y, Z) $ $\sim \prod_{k=1}^K p(x_k) p(y,z| x_{\cal K})$. Then, any rate tuple $\left(R_1^{\text s}, R_1^{\text o}, \cdots,\right.$ $\left. R_K^{\text s}, R_K^{\text o}\right)$ satisfying
	\begin{align}\label{rate_region0}
	\sum_{k \in \cal S} R_k^{\text s} + \sum_{k \in {\cal S}\setminus {\cal S}_1} R_k^{\text o} \leq & \left[I(X_{\cal S}; Y| X_{\bar {\cal S}}) - I(X_{{\cal S}_1}; Z) \right]^+, \nonumber\\
	& \forall {\cal S} \subseteq {\cal K} ~{\text {and}}~ {{\cal S}_1} \subseteq \cal S.
	\end{align}
	is achievable.
	Let ${\mathscr R} (X_{\cal K})$ denote the set of rate tuples satisfying (\ref{rate_region0}).
	Then, the convex hull of the union of ${\mathscr R} (X_{\cal K})$ over all $\prod_{k=1}^K p(x_k)$ is an achievable secrecy rate region of the MAC wiretap channel with $K$ users.
\end{lemma}
\itshape \textbf{Proof:}  \upshape
	This lemma can be proven by a simple extension of the proof of Theorem \ref{theorem1}.
\hfill $\Box$



\begin{remark}\label{remark1}
	In reference \cite{tekin2008general}, the same setting of our paper is considered for the Gaussian MAC wiretap channel.
	A superposition encoding rate region, in which the rate 4-tuples $\left(R_1^{\text s}, R_1^{\text o}, R_2^{\text s}, R_2^{\text o}\right)$ satisfy
	\begin{equation}\label{rate_region11}
	\left\{\!\!\!
	\begin{array}{ll}
	\sum_{k \in {\cal S}} (R_k^{\text s} + R_k^{\text o}) \leq I(X_{\cal S}; Y| X_{\bar {\cal S}}), \forall \cal S \subseteq \cal K, \\
	\sum_{k \in {\cal S}} R_k^{\text s} \leq \left[I(X_{\cal S}; Y| X_{\bar {\cal S}}) - I(X_{\cal S}; Z) \right]^+\!\!, \forall \cal S \subseteq \cal K,
	\end{array} \right.\!\!
	\end{equation}
	is given in \cite[eq.~(19)]{tekin2008general}. Then, it is stated in \cite[Theorem 1]{tekin2008general} that the convex hull of the superposition 
	encoding rate region union over all power constraint is achievable.
	By comparing (\ref{rate_region}) and (\ref{rate_region11}), we notice that they differ in the third inequality of (\ref{rate_region}).
	In Appendix \ref{Appendix_A}, we show that the result in \cite[Theorem 1]{tekin2008general} unfortunately is not correct.
	In this sense, our result provides a general achievable rate region for the MAC-WT scenario with confidential and open messages while
	\cite[Theorem 1]{tekin2008general} does not.
\end{remark}

\section{Proof of Theorem \ref{theorem1}}
\label{section4}


Let $(X_1, X_2) \sim p(x_1) p(x_2)$, and assume that $I(X_{\cal S}; Y| X_{\bar {\cal S}})$ $> I(X_{\cal S}; Z), ~\forall~ \cal S \subseteq \cal K$.
In the following, we show that there exists a $\left( 2^{n R_1^{\text s}}, 2^{n R_1^{\text o}}, 2^{n R_2^{\text s}}, 2^{n R_2^{\text o}}, n \right)$ code such that any rate tuple inside region ${\mathscr R} (X_1, X_2)$, i.e., any $(R_1^{\text s}, R_1^{\text o}, R_2^{\text s}, R_2^{\text o})$ satisfying
\begin{equation}\label{rate_region1}
\left\{\!\!\!
\begin{array}{ll}
\sum_{k \in {\cal S}} (R_k^{\text s} + R_k^{\text o}) < I(X_{\cal S}; Y| X_{\bar {\cal S}}) - \epsilon, \forall \cal S \subseteq \cal K, \\
\sum_{k \in {\cal S}} R_k^{\text s} \!<\! I(X_{\cal S}; Y| X_{\bar {\cal S}}) \!-\! I(X_{\cal S}; Z) \!-\! \epsilon, \!\forall \cal S \subseteq \cal K,\\
R_1^{\text s} + R_2^{\text s} + R_k^{\text o} < I(X_{\cal K}; Y) - I(X_{\bar k}; Z) \!-\! \epsilon, \forall k \in \cal K,
\end{array} \right.\!\!
\end{equation}
is achievable, where $\epsilon$ is an arbitrarily small positive number. This, together with the standard time-sharing over coding strategies, 
suffices to prove the theorem. We start with the following lemma.
\begin{lemma}\label{lemma0}
	For any rate tuple $(R_1^{\text s}, R_1^{\text o}, R_2^{\text s}, R_2^{\text o})$ satisfying (\ref{rate_region1}), there 
	exists a rate pair $(R_1^{\text g}, R_2^{\text g})$ such that
	\begin{equation}\label{rate_region2}
	\left\{\!\!\!
	\begin{array}{ll}
	R_k^{\text g} \geq 0, ~\forall~ k \in \cal K, \\
	\sum_{k \in {\cal S}} (R_k^{\text s} + R_k^{\text o} + R_k^{\text g}) < I(X_{\cal S}; Y| X_{\bar {\cal S}}) - \epsilon, ~\forall~ \cal S \subseteq \cal K, \\
	\sum_{k \in {\cal S}} (R_k^{\text o} + R_k^{\text g}) \geq I(X_{\cal S}; Z), ~\forall~ \cal S \subseteq \cal K,
	\end{array} \right.
	\end{equation}
\end{lemma}
\itshape \textbf{Proof:}  \upshape
	By eliminating $R_1^{\text g}$ and $R_2^{\text g}$ in (\ref{rate_region2}) using the Fourier-Motzkin procedure \cite[Appendix D]{el2011network}, it can be shown that (\ref{rate_region1}) is the projection of (\ref{rate_region2}) onto the hyperplane $\{ R_1^{\text g} = 0, R_2^{\text g} = 0\}$. 
	Lemma \ref{lemma0} can thus be proven.
	Due to space limitation, the detailed procedure is omitted.
\hfill $\Box$

\subsection{Coding Scheme}

For a given rate tuple $(R_1^{\text s}, R_1^{\text o}, R_2^{\text s}, R_2^{\text o})$ inside region ${\mathscr R} (X_1, X_2)$, choose a rate pair $(R_1^{\text g}, R_2^{\text g})$ satisfying (\ref{rate_region2}).
Without loss of generality (w.l.o.g.), assume that $2^{n R_k^{\text s}}$, $2^{n (R_k^{\text o} + R_k^{\text g})}$ and $2^{n R_k^{\text g}}, \forall k \in {\cal K}$ are integers.
Denote
\begin{align}\label{L2}
& {\cal L}_{k,m_k} = \left[ (m_k-1) 2^{n (R_k^{\text o} + R_k^{\text g})} + 1 : m_k 2^{n (R_k^{\text o} + R_k^{\text g})} \right],\nonumber\\
& \quad\quad\quad \forall k \in {\cal K}, m_k \in {\cal M}_k,\nonumber\\
& {\cal L}_k = \left\{ {\cal L}_{k,m_k}, m_k \in {\cal M}_k \right\}\nonumber\\
& \quad\,= \left[ 1:2^{n (R_k^{\text s} + R_k^{\text o} + R_k^{\text g})} \right], ~\forall~ k \in {\cal K}.
\end{align}
Then, a coding scheme is provided below.

{\textbf {Codebook generation.}} 
For each message pair $(m_k, w_k) \in {\cal M}_k \times {\cal W}_k$ of user $k$, generate a subcodebook ${\cal C}_k (m_k)$ by randomly and independently generating $2^{n (R_k^{\text o} + R_k^{\text g})}$ sequences $x_k^n(l_k)$ according to $\prod_{i=1}^n p (x_{ki})$, where $l_k \in {\cal L}_{k,m_k}$. 
These subcodebooks constitute the codebook of user $k$, i.e., ${\cal C}_k = \left\{ {\cal C}_k (m_k), m_k \in {\cal M}_k \right\}$.
The codebooks of all users, i.e., ${\cal C}_k, \forall k \in {\cal K}$, are then revealed to all transmitters and receivers, including the eavesdropper.

{\textbf {Encoding.}} 
Since $R_k^{\text g} \geq 0, \forall k \in \cal K$, evenly divide each subcodebook 
${\cal C}_k(m_k)$ into $2^{n R_k^{\text o}}$ subsets ${\cal C}_k(m_k,w_k)$ of size $2^{n R_k^{\text g}}$ codewords each, for $w_k \in {\cal W}_k$.
To send message pair $(m_k, w_k) \in {\cal M}_k \times {\cal W}_k$, encoder $k$ uniformly chooses a codeword (with index $l_k$) 
from ${\cal C}_k (m_k,w_k)$ and then transmits $x_k^n (l_k)$.

{\textbf {Decoding.}} 
The decoder at the legitimate receiver declares that $({\hat m}_1, {\hat w}_1, {\hat m}_2, {\hat w}_2)$ is sent if it is the unique message tuple such that $(x_1^n (l_1), x_2^n (l_2), y^n) \in {\cal T}_\epsilon^{(n)}(X_1,X_2,Y)$, for some $l_1$ and $l_2$ such that 
$x_k^n (l_k) \in {\cal C}_k ({\hat m}_k,{\hat w}_k)$,  for $ k = 1,2$. 

\subsection{Analysis of the probability of error}

Since $\sum_{k \in {\cal S}} (R_k^{\text s} + R_k^{\text o} + R_k^{\text g}) < I(X_{\cal S}; Y| X_{\bar {\cal S}}) - \epsilon, \forall \cal S \subseteq \cal K$, it can be proven by using LLN and the packing lemma that the probability of error averaged over the random codebook and encoding tends to zero as $n \rightarrow \infty$.
The proof follows exactly the same steps used in \cite[Subsection 4.5.1]{el2011network}.
Hence, $\lim_{n \rightarrow \infty} P_{\text e} \leq \delta$.

\subsection{Analysis of the information leakage rate}

For a given codebook ${\cal C}_k$, the secret message $M_k$ is a function of the codeword index $L_k$.
Hence,
\begin{align}\label{IM1M2Z}
& I(M_1, M_2; Z^n) \nonumber\\
= & H(M_1) + H(M_2) - H(M_1, M_2| Z^n)\nonumber\\
= & n R_1^{\text s} + n R_2^{\text s} - H(L_1, L_2| Z^n) + H(L_1, L_2| M_1, M_2, Z^n).
\end{align}
In order to measure the information leakage rate (\ref{IM1M2Z}), we first transform $H(L_1, L_2| Z^n)$ as follows
\begin{align}\label{HL1L2Z}
& H(L_1, L_2| Z^n) \nonumber\\
= & H(L_1, L_2) - I(L_1, L_2; Z^n)\nonumber\\
\overset{(a)}{=} & H(L_1) + H(L_2) - I(L_1, L_2, X_1^n, X_2^n; Z^n)\nonumber\\
\overset{(b)}{=} & n (R_1^{\text s} + R_1^{\text o} + R_1^{\text g} + R_2^{\text s} + R_2^{\text o} + R_2^{\text g}) - I(X_1^n, X_2^n; Z^n)\nonumber\\
\overset{(c)}{=} & n (R_1^{\text s} + R_1^{\text o} + R_1^{\text g} + R_2^{\text s} + R_2^{\text o} + R_2^{\text g}) - n I(X_1, X_2; Z),
\end{align}
where $(a)$ holds since $X_1^n$ and $X_2^n$ are respectively functions of indexes $L_1$ and $L_2$, $(b)$ holds since $(L_1, L_2) \rightarrow (X_1^n, X_2^n) \rightarrow Z^n$ forms a Markov chain, and $(c)$ follows since $p(x_1^n, x_2^n, z^n) = \prod_{i=1}^n p_{X_1, X_2, Z} (x_{1i}, x_{2i}, z_i)$.
Then, we provide an upper bound on term $H(L_1, L_2| M_1, M_2, Z^n)$ in the following theorem.
\begin{theorem}\label{lemma2}
	\begin{align}\label{upper_bound}
	& \lim_{n \rightarrow \infty} \frac{1}{n} H(L_1, L_2| M_1, M_2, Z^n)\nonumber\\
	& \leq R_1^{\text o} + R_1^{\text g} + R_2^{\text o} + R_2^{\text g} - I(X_1, X_2; Z) + \delta.
	\end{align}
\end{theorem}

\itshape \textbf{Proof:}  \upshape
	See Appendix \ref{Appendix_B}.
\hfill $\Box$

Substituting (\ref{HL1L2Z}) and (\ref{upper_bound}) into (\ref{IM1M2Z}), we have
\begin{equation}\label{I_leq_delta}
\lim_{n \rightarrow \infty} R_{{\text E}, {\cal K}} = \lim_{n \rightarrow \infty} \frac{1}{n} I(M_1, M_2; Z^n) \leq \delta.
\end{equation}

Theorem \ref{theorem1} is thus proven.

\section{Conclusions}
\label{section6}

In this paper, we studied the capacity region of a discrete memoryless (DM) multiple access wiretap (MAC-WT) channel where,
besides confidential messages,  the users have also open messages to transmit.
Different from \cite{tekin2008general}, which assumed Gaussian inputs and Gaussian channels, we considered general inputs and DM channels.
By using random coding, we found an achievable rate region where the information leakage of the confidential messages to the eavesdropper 
and the probability of error of all messages at the intended receiver vanish as the block length increase to infinity. 
Furthermore, we also correct the result in \cite{tekin2008general} that studied the same scenario in the Gaussian MAC case, 
but where the provided achievable region is actually not generally achievable.

\appendices

\section{}
\label{Appendix_A}

For brevity, we also consider the two-user case for reference \cite{tekin2008general}.
When proving \cite[Theorem 1]{tekin2008general}, it is stated in \cite{tekin2008general} that for any rate tuple $\left(R_1^{\text s}, R_1^{\text o}, R_2^{\text s}, R_2^{\text o}\right)$ satisfying (\ref{rate_region11}), there exists $R_k^x$ such that \cite[eq. (26) -- (28)]{tekin2008general} hold.
When $R_k^{\text o}$ is large, to ensure that \cite[eq. (27)]{tekin2008general} is satisfied, some open message of user $k$ can be reclassified as secret message (we call this rate splitting in the following).
For the sake of convenience, we rewrite \cite[eq. (26) -- (28)]{tekin2008general} as follows
\begin{align}
& \sum_{k \in {\cal S}} (R_k^{\text s} + R_k^{\text o} + R_k^x) \leq I(X_{\cal S}; Y| X_{\bar {\cal S}}), ~\forall~ \cal S \subseteq \cal K, \label{rate_region20_a}\\
& \sum_{k \in {\cal S}} (R_k^{\text o} + R_k^x) \leq I(X_{\cal S}; Z| X_{\bar {\cal S}}), \forall \cal S \subseteq \cal K,\nonumber\\
&\quad\quad\quad\quad\quad\quad\quad\quad\quad\quad\quad {\text {with equality if }} \cal S = \cal K, \label{rate_region20_b}\\
& \sum_{k \in {\cal S}} R_k^{\text s} \leq \left[I(X_{\cal S}; Y| X_{\bar {\cal S}}) - I(X_{\cal S}; Z)\right]^+, ~\forall~ \cal S \subseteq \cal K. \label{rate_region20_c}
\end{align}
Note that for comparison, we replace $\log$ expressions for the Gaussian case with Gaussian inputs in \cite{tekin2008general} with mutual informations.
Moreover, although not mentioned, it is clear by the definition of $R_k^x$ that
\begin{equation}\label{Rxk}
R_k^x \geq 0, ~\forall~ k \in \cal K.
\end{equation}

In order to check whether it is true that for any rate tuple $\left(R_1^{\text s}, R_1^{\text o}, R_2^{\text s}, R_2^{\text o}\right)$ satisfying (\ref{rate_region11}), with rate splitting, there exists $R_k^x$ such that (\ref{rate_region20_a}) -- (\ref{Rxk}) hold, 
we eliminate $R_k^x$ in (\ref{rate_region20_a}) -- (\ref{Rxk}) using the Fourier-Motzkin procedure \cite[Appendix D]{el2011network}, and get
\begin{equation}\label{rate_region21}
\left\{\!\!\!
\begin{array}{ll}
\sum_{k \in {\cal S}} (R_k^{\text s} + R_k^{\text o}) \leq I(X_{\cal S}; Y| X_{\bar {\cal S}}), \forall \cal S \subseteq \cal K, \\
\sum_{k \in {\cal S}} R_k^{\text s} \leq \left[I(X_{\cal S}; Y| X_{\bar {\cal S}}) - I(X_{\cal S}; Z) \right]^+\!\!, \forall \cal S \subseteq \cal K,\\
\sum_{k \in {\cal S}} R_k^{\text o} \leq I(X_{\cal S}; Z| X_{\bar {\cal S}}), \forall \cal S \subseteq \cal K.
\end{array} \right.\!\!
\end{equation}
Denote the sets of rate tuples $(R_1^{\text s}, R_1^{\text o}, R_2^{\text s}, R_2^{\text o})$ satisfying (\ref{rate_region}), (\ref{rate_region11}) and (\ref{rate_region21}) by $\mathscr R$, ${\mathscr R}_1$ and ${\mathscr R}_2$, respectively.
Then, if \cite[Theorem 1]{tekin2008general} is true, with rate splitting, all rate tuples in region ${\mathscr R}_1$ should be able to be transformed to rate tuples in region ${\mathscr R}_2$.
However, in the following we show that ${\mathscr R}_2$ is equivalent to $\mathscr R$, and there exist rate tuples in region ${\mathscr R}_1$ which can not be transformed to rate tuples in region ${\mathscr R}_2$.

For any given rate tuple $A  = (R_1^{\text s}, R_1^{\text o}, R_2^{\text s}, R_2^{\text o})$ in region ${\mathscr R}_2$, it is obvious from (\ref{rate_region21}) that 
\begin{equation}
\left\{\!\!\!
\begin{array}{ll}
R_1^{\text s} + R_1^{\text o} + R_2^{\text s} \leq \left[I(X_1, X_2; Y) - I(X_2; Z) \right]^+, \\
R_1^{\text s} + R_2^{\text s} + R_2^{\text o} \leq \left[I(X_1, X_2; Y) - I(X_1; Z) \right]^+.
\end{array} \right.\!\!
\end{equation}
Hence, $A$ is also in region $\mathscr R$.

\begin{figure}
	\centering
	\includegraphics[width=6cm]{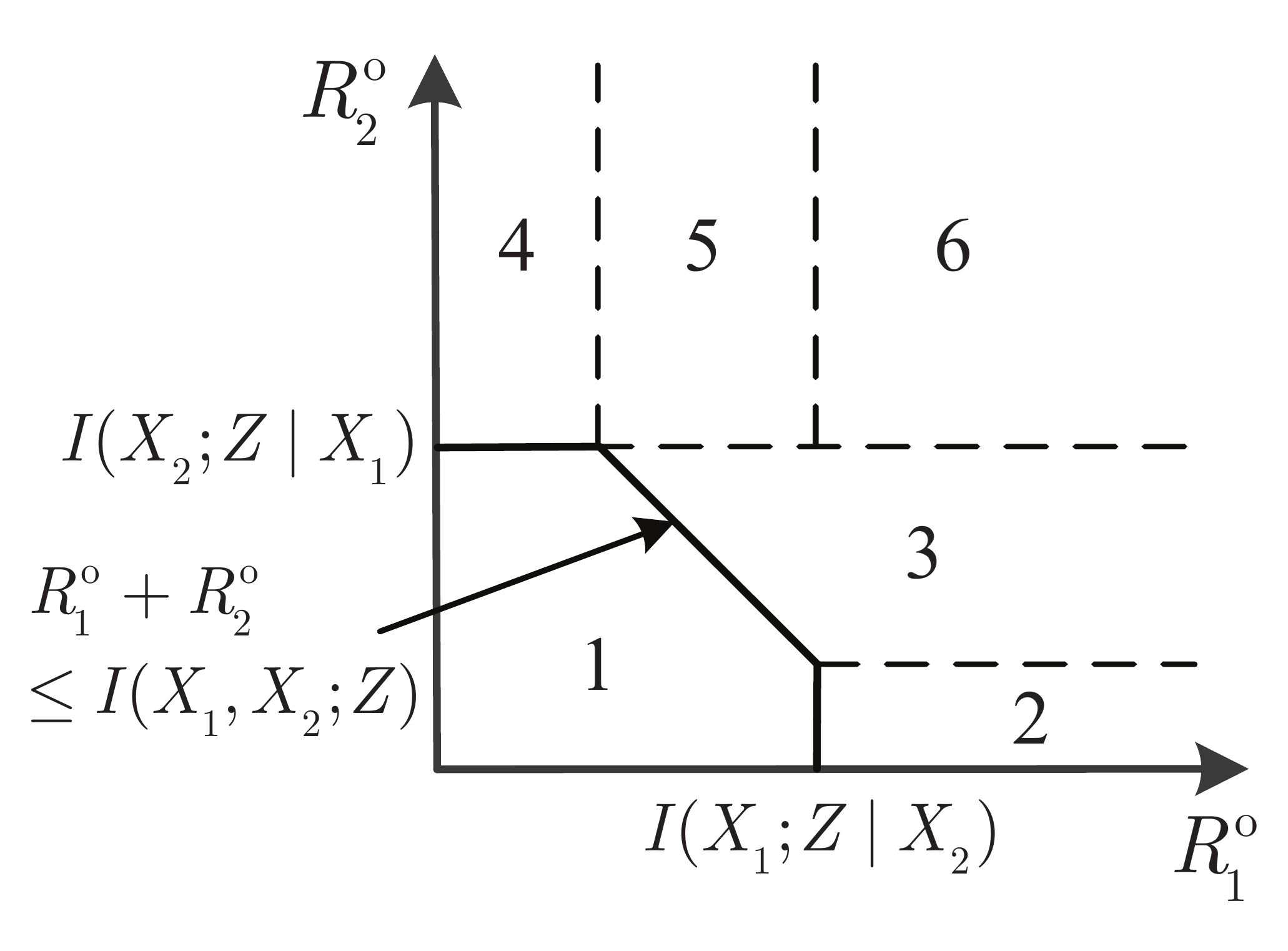}
	\caption{Classification of rate tuples.}
	\label{Fig2}
\end{figure}

Based on the values of $R_1^{\text o}$ and $R_2^{\text o}$, all rate tuples in region $\mathscr R$ can be divided into $6$ categories as shown in Fig. \ref{Fig2}. 
In the following we show that, with rate splitting, any given rate tuple $B = (R_1^{\text s}, R_1^{\text o}, R_2^{\text s}, R_2^{\text o})$ in region $\mathscr R$ can be transformed to another rate tuple $B'$ in region $\mathscr R_2$.

If rate tuple $B$ belongs to category $1$, i.e.,
\begin{equation}\label{category1}
\sum_{k \in {\cal S}} R_k^{\text o} \leq I(X_{\cal S}; Z| X_{\bar {\cal S}}), \forall \cal S \subseteq \cal K,
\end{equation}
it is obvious that $B$ is also in region $\mathscr R_2$.

If rate tuple $B$ belongs to category $2$, i.e.,
\begin{equation}\label{category2}
\left\{\!\!\!
\begin{array}{ll}
R_1^{\text o} > I(X_1; Y| X_2), \\
0 \leq R_2^{\text o} \leq I(X_2; Z),
\end{array} \right.
\end{equation}
let
\begin{align}\label{category2_trans}
& {\tilde R}_1^{\text o} = I(X_1; Z| X_2), \nonumber\\
& {\tilde R}_1^{\text s} = R_1^{\text s} + R_1^{\text o} - I(X_1; Z| X_2).
\end{align}
We get a new rate tuple $B' = ({\tilde R}_1^{\text s}, {\tilde R}_1^{\text o}, R_2^{\text s}, R_2^{\text o})$.
Since $B$ is in region $\mathscr R$, it satisfies (\ref{rate_region}).
Hence, 
\begin{align}
& {\tilde R}_1^{\text o} + R_2^{\text o} = I(X_1; Z| X_2) + R_2^{\text o}\nonumber\\
& \quad\quad\quad\quad\!\! \leq I(X_1, X_2; Z),\nonumber\\
& {\tilde R}_1^{\text s} + {\tilde R}_1^{\text o} = R_1^{\text s} + R_1^{\text o}\nonumber\\
& \quad\quad\quad\quad\!\! \leq I(X_1; Y| X_2), \nonumber\\
& {\tilde R}_1^{\text s} + {\tilde R}_1^{\text o} + R_2^{\text s} + R_2^{\text o} = R_1^{\text s} + R_1^{\text o} + R_2^{\text s} + R_2^{\text o}\nonumber\\
& \quad\quad\quad\quad\quad\quad\quad\quad\quad\!\!\! \leq I(X_1, X_2; Y), \nonumber\\
& {\tilde R}_1^{\text s} = R_1^{\text s} + R_1^{\text o} - I(X_1; Y| X_2)\nonumber\\
& \quad \, \leq I(X_1; Y| X_2) - I(X_1; Z| X_2)\nonumber\\
& \quad \, \leq \left[ I(X_1; Y| X_2) - I(X_1; Z) \right]^+,\nonumber\\
& {\tilde R}_1^{\text s} + R_2^{\text s} = R_1^{\text s} + R_1^{\text o} + R_2^{\text s} - I(X_1; Y| X_2) \nonumber\\
& \quad\quad\quad\quad\!\! \leq \left[ I(X_1, X_2; Y) - I(X_2; Z) \right]^+ - I(X_1; Z| X_2)\nonumber\\
& \quad\quad\quad\quad\!\! \leq \left[ I(X_1, X_2; Y) - I(X_1, X_2; Z) \right]^+.
\end{align}
The values of $R_2^{\text s} + R_2^{\text o}$ and $R_2^{\text s}$ remain unchanged.
Hence, $B'$ is in region $\mathscr R_2$.

If $B$ belongs to category $3$, i.e.,
\begin{equation}\label{category3}
\left\{\!\!\!
\begin{array}{ll}
I(X_2; Z) < R_2^{\text o} \leq I(X_2; Z| X_1), \\
R_1^{\text o} + R_2^{\text o} > I(X_1, X_2; Z),
\end{array} \right.
\end{equation}
let
\begin{align}\label{category3_trans}
& {\tilde R}_1^{\text o} = I(X_1, X_2; Z) - R_2^{\text o}, \nonumber\\
& {\tilde R}_1^{\text s} = R_1^{\text s} + R_1^{\text o} + R_2^{\text o} - I(X_1, X_2; Z).
\end{align}
A new rate tuple $B' = ({\tilde R}_1^{\text s}, {\tilde R}_1^{\text o}, R_2^{\text s}, R_2^{\text o})$ is then obtained.
Since  
\begin{align}
& {\tilde R}_1^{\text o} + R_2^{\text o} = I(X_1, X_2; Z),\nonumber\\
& {\tilde R}_1^{\text s} + {\tilde R}_1^{\text o} = R_1^{\text s} + R_1^{\text o}\nonumber\\
& \quad\quad\quad\quad\!\! \leq I(X_1; Y| X_2), \nonumber\\
& {\tilde R}_1^{\text s} + {\tilde R}_1^{\text o} + R_2^{\text s} + R_2^{\text o} = R_1^{\text s} + R_1^{\text o} + R_2^{\text s} + R_2^{\text o}\nonumber\\
& \quad\quad\quad\quad\quad\quad\quad\quad\quad\!\!\! \leq I(X_1, X_2; Y), \nonumber\\
& {\tilde R}_1^{\text s} = R_1^{\text s} + R_1^{\text o} + R_2^{\text o} - I(X_1, X_2; Z)\nonumber\\
& \quad\, \leq I(X_1; Y| X_2) + R_2^{\text o} - I(X_1; Z) - I(X_2; Z| X_1)\nonumber\\
& \quad\, \leq \left[ I(X_1; Y| X_2) - I(X_1; Z) \right]^+,\nonumber\\
& {\tilde R}_1^{\text s} + R_2^{\text s} = R_1^{\text s} + R_1^{\text o} + R_2^{\text s} + R_2^{\text o} - I(X_1, X_2; Z) \nonumber\\
& \quad\quad\quad\quad\!\! \leq \left[ I(X_1, X_2; Y) - I(X_1, X_2; Z) \right]^+,
\end{align}
and the values of $R_2^{\text s} + R_2^{\text o}$ and $R_2^{\text s}$ remain unchanged, $B'$ is in region $\mathscr R_2$.

Analogously, if $B$ belongs to category $4$, i.e.,
\begin{equation}\label{category4}
\left\{\!\!\!
\begin{array}{ll}
0 \leq R_1^{\text o} \leq I(X_1; Z), \\
R_2^{\text o} > I(X_2; Z| X_1),
\end{array} \right.
\end{equation}
let
\begin{align}\label{category4_trans}
& {\tilde R}_2^{\text o} = I(X_2; Z| X_1), \nonumber\\
& {\tilde R}_2^{\text s} = R_2^{\text s} + R_2^{\text o} - I(X_2; Z| X_1),
\end{align}
and if $B$ belongs to category $5$, i.e.,
\begin{equation}\label{category5}
\left\{\!\!\!
\begin{array}{ll}
I(X_1; Z) < R_1^{\text o} \leq I(X_1; Z| X_2), \\
R_2^{\text o} > I(X_2; Z| X_1),
\end{array} \right.
\end{equation}
let
\begin{align}\label{category5_trans}
& {\tilde R}_2^{\text o} = I(X_1, X_2; Z) - R_1^{\text o}, \nonumber\\
& {\tilde R}_2^{\text s} = R_1^{\text o} + R_2^{\text s} + R_2^{\text o} - I(X_1, X_2; Z).
\end{align}
It can be similarly proven that the newly obtained rate tuple $B' = (R_1^{\text s}, R_1^{\text o}, {\tilde R}_2^{\text s}, {\tilde R}_2^{\text o})$ is in region $\mathscr R_2$.

If $B$ belongs to category $6$, i.e.,
\begin{equation}\label{category6}
\left\{\!\!\!
\begin{array}{ll}
R_1^{\text o} > I(X_1; Z| X_2), \\
R_2^{\text o} > I(X_2; Z| X_1),
\end{array} \right.
\end{equation}
let
\begin{align}\label{category6_trans}
& {\tilde R}_1^{\text o} = I(X_1; Z| X_2), \nonumber\\
& {\tilde R}_1^{\text s} = R_1^{\text s} + R_1^{\text o} - I(X_1; Z| X_2),\nonumber\\
& {\tilde R}_2^{\text o} = I(X_2; Z), \nonumber\\
& {\tilde R}_2^{\text s} = R_2^{\text s} + R_2^{\text o} - I(X_2; Z).
\end{align}
Then, 
\begin{align}\label{category6_trans1}
& {\tilde R}_1^{\text o} + {\tilde R}_2^{\text o} = I(X_1, X_2; Z), \nonumber\\
& \sum_{k \in {\cal S}} ({\tilde R}_k^{\text s} + {\tilde R}_k^{\text o}) =\sum_{k \in {\cal S}} (R_k^{\text s} + R_k^{\text o})\nonumber\\
& \quad\quad\quad\quad\quad\quad\, \leq I(X_{\cal S}; Y| X_{\bar {\cal S}}), \forall \cal S \subseteq \cal K,\nonumber\\
& {\tilde R}_1^{\text s} \leq I(X_1; Y| X_2) - I(X_1; Z| X_2)\nonumber\\
& \quad\, \leq \left[ I(X_1; Y| X_2) - I(X_1; Z) \right]^+, \nonumber\\
& {\tilde R}_2^{\text s} \leq \left[ I(X_2; Y| X_1) - I(X_2; Z) \right]^+, \nonumber\\
& {\tilde R}_1^{\text s} + {\tilde R}_2^{\text s} = R_1^{\text s} + R_1^{\text o} + R_2^{\text s} + R_2^{\text o} - I(X_1, X_2; Z)\nonumber\\
& \quad\quad\quad\quad \!\!\leq \left[ I(X_1, X_2; Y) - I(X_1, X_2; Z) \right]^+.
\end{align}
Rate tuple $B' =  ({\tilde R}_1^{\text s}, {\tilde R}_1^{\text o}, {\tilde R}_2^{\text s}, {\tilde R}_2^{\text o})$ is thus in region $\mathscr R_2$.

Until now, we have shown that any rate tuple $A$ in region $\mathscr R_2$ is also in region $\mathscr R$, and by using rate splitting, any rate tuple $B$ in region $\mathscr R$ can be transformed to another rate tuple $B'$ in region $\mathscr R_2$.
Therefore, with rate splitting ${\mathscr R}_2$ is equivalent to $\mathscr R$.

Next, we show that there exist rate tuples in region ${\mathscr R}_1$ which can not be transformed to rate tuples in region ${\mathscr R}_2$.
Consider rate tuple $C = (R_1^{\text s}, R_1^{\text o}, R_2^{\text s}, R_2^{\text o})$, where
\begin{align}\label{Counterexample}
& R_1^{\text s} = \left[I(X_1; Y| X_2) - I(X_1; Z)\right]^+, \nonumber\\
& R_1^{\text o} = 0,\nonumber\\
& R_2^{\text s} = \left[I(X_2; Y) - I(X_2; Z| X_1)\right]^+, \nonumber\\
& R_2^{\text o} = I(X_1, X_2; Z).
\end{align}
Since $(X_1, X_2, Y, Z) \!\sim~\!p(x_1) p(x_2) p(y,z| x_1, x_2)$, it is possible that 
\begin{align}
& I(X_2; Y) + I(X_1; Z) \leq I(X_2; Y| X_1),\nonumber\\
& I(X_1, X_2; Z) > I(X_2; Z| X_1).
\end{align}
When the above inequality holds, it can be easily found that rate tuple $C$ satisfies (\ref{rate_region11}) and is thus in region ${\mathscr R}_1$.
However, since $R_2^{\text o} > I(X_2; Z| X_1)$, (\ref{rate_region21}) is not satisfied, and $C$ is thus outside region ${\mathscr R}_2$.
Because $R_1^{\text s} + R_2^{\text s} = \left[I(X_1, X_2; Y) - I(X_1, X_2; Z)\right]^+$, it would be impossible to reduce $R_2^{\text o}$ by increasing $R_2^{\text s}$, i.e., reclassifying some open message of user $2$ as secret message of user $2$, since otherwise $R_1^{\text s} + R_2^{\text s} > \left[I(X_1, X_2; Y) - I(X_1, X_2; Z)\right]^+$.
In this case, rate tuple $C$ can not be transformed to another rate tuple in region ${\mathscr R}_1$, indicating that not all rate tuples in region ${\mathscr R}_1$ can be transformed to rate tuples in region ${\mathscr R}_2$ even with rate splitting.

Based on the above analysis, it can be concluded that the statement of \cite[Theorem 1]{tekin2008general} doe snot hold in general.
 
\section{Proof of Theorem \ref{lemma2}}
\label{Appendix_B}

For given $n$-th order product distribution on ${\cal X}^n_1 \times {\cal X}^n_2 \times {\cal Z}^n$, 
recall the definition of conditional $\epsilon$-typical sets
	\begin{align}
	&\!{\cal T}_\epsilon^{(n)} (X_1, X_2| z^n)\nonumber\\
	&\! = \left\{ (x_1^n, x_2^n)| (x_1^n, x_2^n, z^n) \in {\cal T}_\epsilon^{(n)} (X_1, X_2, Z) \right\},\label{T_conditional}\\
	& \!{\cal T}_\epsilon^{(n)}\! (X_2| x_1^n, z^n) \!=\! \left\{\! x_2^n| (x_1^n, x_2^n, z^n) \!\in\! {\cal T}_\epsilon^{(n)}\! (X_1, X_2, Z) \!\right\}\!,\!\!\! \label{T_conditional2}\\
	& \!{\cal T}_\epsilon^{(n)}\! (X_1| x_2^n, z^n) \!=\! \left\{\! x_1^n| (x_1^n, x_2^n, z^n) \!\in\! {\cal T}_\epsilon^{(n)}\! (X_1, X_2, Z) \!\right\}\!.\!\!\! \label{T_conditional3}
	\end{align}
To prove Theorem \ref{lemma2}, we bound $H(L_1, L_2| m_1, m_2, Z^n)$ for every secret message pair $(m_1, m_2)$.
First, for a given received signal $z^n$ at the eavesdropper, assume that it is a typical sequence, i.e., $z^n \in {\cal T}_\epsilon^{(n)} (Z)$, and define
\begin{align}\label{D}
{\cal D} (m_1, m_2, z^n) \!=\!& \left\{\! (l_1, l_2)| (x_1^n (l_1), x_2^n (l_2)) \!\in\! {\cal T}_\epsilon^{(n)} (X_1, X_2| z^n), \right.\nonumber\\
& \left. \forall~ (l_1, l_2) \in {\cal L}_{1,m_1} \times {\cal L}_{2,m_2} \right\},
\end{align}
and
\begin{equation}\label{N}
N(m_1, m_2, z^n) = \left|{\cal D} (m_1, m_2, z^n)\right|.
\end{equation}
In the following theorem, we give an upper bound on the expectation and the variance of $N(m_1, m_2, z^n)$.
\begin{theorem}\label{theorem3}
The expectation and variance of $N(m_1, m_2, z^n)$ can be bounded as
\begin{align}
{\mathbb E} \left\{ N(m_1, m_2, z^n) \right\} & \leq 2^{n (\Delta + \delta_1 (\epsilon))}, \label{EN}\\
{\text {Var}} \left\{ N(m_1, m_2, z^n) \right\} & \leq 2^{n (\Delta + \delta_1 (\epsilon))} + \sum_{k \in \cal K} 2^{n (2 \Delta - \Delta_k + \delta_1 (\epsilon))},\label{VarN}
\end{align}
where $\delta_1 (\epsilon)$ is given in (\ref{p_1_up}), and
\begin{align}\label{delta}
& \Delta = R_1^{\text o} + R_1^{\text g} + R_2^{\text o} + R_2^{\text g} - I(X_1, X_2; Z),\nonumber\\
& \Delta_k = R_k^{\text o} + R_k^{\text g} - I(X_k; Z), ~\forall~ k \in {\cal K}.
\end{align}
\end{theorem}

\itshape \textbf{Proof:}  \upshape
	See Appendix \ref{Appendix_C}.
\hfill $\Box$

Next, define the event 
\begin{equation}
{\cal E} (m_1, m_2, z^n) = \left\{  N(m_1, m_2, z^n) \geq 2^{n (\Delta + \delta_1 (\epsilon)) + 1} \right\}.
\end{equation}
We have
\begin{align}\label{p_event}
& P\left\{{\cal E} (m_1, m_2, z^n)\right\} \nonumber\\
= & P \left\{  N(m_1, m_2, z^n) \geq 2^{n (\Delta + \delta_1 (\epsilon)) + 1} \right\}\nonumber\\
\leq & P \left\{  N(m_1, m_2, z^n) \geq {\mathbb E} \left\{ N(m_1, m_2, z^n) \right\} + 2^{n (\Delta + \delta_1 (\epsilon))} \right\}\nonumber\\
\leq & P \left\{ \left| N(m_1, m_2, z^n) \!-\! {\mathbb E} \left\{ N(m_1, m_2, z^n) \right\} \right| \!\geq\! 2^{n (\Delta + \delta_1 (\epsilon))} \right\}\nonumber\\
\overset{(a)}{\leq} & \frac{{\text {Var}} \left\{ N(m_1, m_2, z^n) \right\}}{2^{2n (\Delta + \delta_1 (\epsilon))}}\nonumber\\
\overset{(b)}{\leq} & 2^{- n (\Delta + \delta_1 (\epsilon))} + \sum_{k \in \cal K} 2^{ - n (\Delta_k + \delta_1 (\epsilon))},
\end{align}
where step $(a)$ follows by applying the Chebyshev inequality, and $(b)$ follows by (\ref{VarN}).
Due to (\ref{rate_region2}), $\Delta \geq 0$ and $\Delta_k \geq 0, \forall k \in {\cal K}$.
Then, it is obvious that $P\left\{{\cal E} (m_1, m_2, z^n)\right\} \rightarrow 0$ as $n \rightarrow \infty$.
For any $z^n \in {\cal T}_\epsilon^{(n)} (Z)$, define indicator variable
\begin{equation}\label{indicator_E}
E (m_1, m_2, z^n) \!=\! \left\{\!\!\!
\begin{array}{ll}
1,& \!\! {\text {if}}~ {\cal E} (m_1, m_2, z^n) ~{\text {occurs}},\\
0,& \!\! {\text {otherwise}}.\\
\end{array} \right.
\end{equation} 
Then, $P \left\{ E (m_1, m_2, z^n) = 1 \right\} \rightarrow 0$ as $n \rightarrow \infty$.

\newcounter{TempEqCnt}
\setcounter{TempEqCnt}{\value{equation}}
\setcounter{equation}{65}
\begin{figure*}[ht]
	\begin{align}\label{N_square}
	& {\mathbb E} \left\{[N(m_1, m_2, z^n)]^2\right\} = \sum_{(l_1,l_2) \in {\cal L}_{1,m_1} \times {\cal L}_{2,m_2}} \left\{ P \left\{ (x_1^n (l_1), x_2^n (l_2)) \in {\cal T}_\epsilon^{(n)} (X_1, X_2| z^n) \right\} \right.\nonumber\\
	& + \sum_{l_2' \in {\cal L}_{2,m_2}\setminus \{l_2\} } P \left\{ (x_1^n (l_1), x_2^n (l_2)) \in {\cal T}_\epsilon^{(n)} (X_1, X_2| z^n),~ x_2^n (l_2') \in {\cal T}_\epsilon^{(n)} (X_2| x_1^n  (l_1), z^n) \right\}\nonumber\\
	& + \sum_{l_1' \in {\cal L}_{1,m_1}\setminus \{l_1\} } P \left\{ (x_1^n (l_1), x_2^n (l_2)) \in {\cal T}_\epsilon^{(n)} (X_1, X_2| z^n),~ x_1^n (l_1') \in {\cal T}_\epsilon^{(n)} (X_1| x_2^n  (l_2), z^n) \right\}\nonumber\\
	& + \sum_{(l_1',l_2') \in {\cal L}_{1,m_1} \times {\cal L}_{2,m_2} \setminus \left\{(l_1, l_2)\right\} } \left. P \left\{ (x_1^n (l_1), x_2^n (l_2)), (x_1^n (l_1'), x_2^n (l_2')) \in {\cal T}_\epsilon^{(n)} (X_1, X_2| z^n) \right\} \right\}\nonumber\\
	& = 2^{n (R_1^{\text o} + R_1^{\text g} + R_2^{\text o} + R_2^{\text g})} \left\{ p_1 + (2^{n (R_2^{\text o} + R_2^{\text g})} - 1) p_2 + (2^{n (R_1^{\text o} + R_1^{\text g})} - 1) p_3 + (2^{n(R_1^{\text o} + R_1^{\text g} + R_2^{\text o} + R_2^{\text g})} - 1) p_4 \right\}\nonumber\\
	& \leq 2^{n (\Delta + \delta_1 (\epsilon))} + \sum_{k \in \cal K} 2^{n (2 \Delta - \Delta_k + \delta_1 (\epsilon))} + \left\{{\mathbb E} [N(m_1, m_2, z^n)]\right\}^2.
	\end{align}
	\hrulefill
\end{figure*}

Since there are $2^{n(R_k^{\text o} + R_k^{\text g})}$ codewords in each subcodebook ${\cal C}_k (m_k), \forall k \in \cal K$, we have
\setcounter{equation}{48}
\begin{align}\label{HL1L2Zm1m21}
& H(L_1, L_2| m_1, m_2, z^n) \nonumber\\
\leq & \log (2^{n(R_1^{\text o} + R_1^{\text g} + R_2^{\text o} + R_2^{\text g})})\nonumber\\
= & n(R_1^{\text o} + R_1^{\text g} + R_2^{\text o} + R_2^{\text g}), ~\forall~ z^n \in {\cal Z}^n,
\end{align}
and 
\begin{align}\label{HL1L2Zm1m23}
& H(L_1, L_2| m_1, m_2, Z^n) \nonumber\\
= & \sum_{z^n \in {\cal Z}^n} p(z^n) H(L_1, L_2| m_1, m_2, z^n)\nonumber\\
\leq & n(R_1^{\text o} + R_1^{\text g} + R_2^{\text o} + R_2^{\text g}).
\end{align}
Moreover, based on the definition of $N(m_1, m_2, z^n)$ in (\ref{N}), we have
\begin{align}\label{HL1L2Zm1m22}
& H(L_1, L_2| m_1, m_2, E (m_1, m_2, z^n) = 0, z^n)\nonumber\\
\leq & \log(N(m_1, m_2, z^n))\nonumber\\
\leq & n (\Delta + \delta_1 (\epsilon)) + 1, ~\forall~ z^n \in {\cal T}_\epsilon^{(n)} (Z),
\end{align}
where the last step holds due to the fact that when $E (m_1, m_2, z^n) = 0$, $N(m_1, m_2, z^n) \leq 2^{n (\Delta + \delta_1 (\epsilon)) + 1}$.
Based on (\ref{HL1L2Zm1m21}), (\ref{HL1L2Zm1m23}) and (\ref{HL1L2Zm1m22}), $H(L_1, L_2| m_1, m_2, Z^n)$ can be upper-bounded as follows
\begin{align}\label{HL1L2Zm1m2}
& H(L_1, L_2| m_1, m_2, Z^n)\nonumber\\
= & P\left\{ Z^n \!\in\! {\cal T}_\epsilon^{(n)} (Z) \right\} H(L_1, L_2| m_1, m_2, Z^n, Z^n \!\in\! {\cal T}_\epsilon^{(n)} (Z))\nonumber\\
+ & P\left\{ Z^n \!\notin\! {\cal T}_\epsilon^{(n)} (Z) \right\} H(L_1, L_2| m_1, m_2, Z^n, Z^n \!\notin\! {\cal T}_\epsilon^{(n)} (Z))\nonumber\\
\leq & \sum_{z^n \in {\cal T}_\epsilon^{(n)} (Z)} p (z^n) H(L_1, L_2| m_1, m_2, z^n) + n \alpha_1\nonumber\\
= & \!\!\sum_{z^n \in {\cal T}_\epsilon^{(n)} (Z)}\!\!\!\!\! \left\{  p_1 (z^n) H(L_1, L_2| m_1, m_2, E (m_1, m_2, z^n) \!=\! 1, z^n) \right.\nonumber\\
+ & \left.  p_2 (z^n) H(L_1, L_2| m_1, m_2, E (m_1, m_2, z^n) \!=\! 0, z^n)\right\} + n \alpha_1\nonumber\\
\leq & \sum_{z^n \in {\cal T}_\epsilon^{(n)} (Z)} \big \{ p (z^n) \alpha_2 H(L_1, L_2| m_1, m_2, z^n) \nonumber\\
+ & p (z^n) H(L_1, L_2| m_1, m_2, E (m_1, m_2, z^n) = 0, z^n) \big \} + n \alpha_1\nonumber\\
\leq & n (\Delta + \delta_2 (\epsilon)),
\end{align}
where
\begin{align}\label{del}
& \alpha_1 = P\left\{ Z^n \notin {\cal T}_\epsilon^{(n)} (Z) \right\} (R_1^{\text o} + R_1^{\text g} + R_2^{\text o} + R_2^{\text g}),\nonumber\\
& p_1 (z^n) = p (z^n) P \left\{ E (m_1, m_2, z^n) = 1 \right\}, ~\forall~ z^n \in {\cal T}_\epsilon^{(n)} (Z)\nonumber\\
& p_2 (z^n) = p (z^n) P \left\{ E (m_1, m_2, z^n) = 0 \right\}, ~\forall~ z^n \in {\cal T}_\epsilon^{(n)} (Z)\nonumber\\
& \alpha_2 = \max \left\{ P \left\{ E (m_1, m_2, z^n) = 1 \right\}, ~\forall~ z^n \in {\cal T}_\epsilon^{(n)} (Z) \right\},\nonumber\\
& \delta_2 (\epsilon) = \delta_1 (\epsilon) + \alpha_2 (R_1^{\text o} + R_1^{\text g} + R_2^{\text o} + R_2^{\text g}) + \frac{1}{n} + \alpha_1.
\end{align}
By the LLN, $P\left\{ Z^n \notin {\cal T}_\epsilon^{(n)} (Z) \right\} \rightarrow 0$ as $n \rightarrow \infty$. 
Hence, $\alpha_1 \rightarrow 0$ as $n \rightarrow \infty$.
In addition, since $P \left\{ E (m_1, m_2, z^n) = 1 \right\} \rightarrow 0, \forall z^n \in {\cal T}_\epsilon^{(n)} (Z)$ as $n \rightarrow \infty$, $\alpha_2 \rightarrow 0$ as $n \rightarrow \infty$. 
$\delta_2 (\epsilon)$ can thus be arbitrarily small as $n \rightarrow \infty$.
Hence,
\begin{align}\label{upper_bound1}
& \lim_{n \rightarrow \infty} \frac{1}{n} H(L_1, L_2| M_1, M_2, Z^n)\nonumber\\
& = \lim_{n \rightarrow \infty} \sum_{m_1 = 1}^{2^{n R_1^{\text s}}} \sum_{m_2 = 1}^{2^{n R_2^{\text s}}} \frac{1}{n} 2^{-n (R_1^{\text s} + R_2^{\text s})} H(L_1, L_2| m_1, m_2, Z^n)\nonumber\\
&\leq \Delta + \delta.
\end{align}
Theorem \ref{lemma2} is thus proven.

\section{Proof of Theorem \ref{theorem3}}
\label{Appendix_C}

Using the conditional typicality lemma, for sufficiently large $n$, we have
\begin{align}
& |{\cal T}_\epsilon^{(n)} (X_1, X_2| z^n)| \leq 2^{n(H(X_1, X_2| Z) + \epsilon)}, \label{T_conditional1}\\
& |{\cal T}_\epsilon^{(n)} (X_2| x_1^n, z^n)| \leq 2^{n(H(X_2| X_1, Z) + \epsilon)},\label{T_conditional4}\\
& |{\cal T}_\epsilon^{(n)} (X_1| x_2^n, z^n)| \leq 2^{n(H(X_1| X_2, Z) + \epsilon)}.\label{T_conditional5}
\end{align}
Let 
\begin{equation}\label{p_1}
p_1 = P \left\{ (X_1^n, X_2^n) \in {\cal T}_\epsilon^{(n)} (X_1, X_2| z^n) \right\}.
\end{equation}
Since $X_1^n$ and $X_2^n$ are independent, an upper bound of $p_1$ can be obtained as follows
\begin{align}\label{p_1_up}
p_1 & = \sum_{(x_1^n, x_2^n) \in {\cal T}_\epsilon^{(n)} (X_1, X_2| z^n)} p(x_1^n) p(x_2^n)\nonumber\\
& \leq 2^{n(H(X_1, X_2| Z) + \epsilon)} 2^{-n(H(X_1) - \epsilon)} 2^{-n(H(X_2) - \epsilon)}\nonumber\\
& \leq 2^{-n(I(X_1, X_2; Z) - \delta_1 (\epsilon))},
\end{align}
where $\delta_1 (\epsilon) = 5 \epsilon$.
Furthermore, denote
\begin{align}\label{p2p3p4}
& p_2 \!=\! P \left\{\! (X_1^n, X_2^n) \!\in\! {\cal T}_\epsilon^{(n)} (X_1, X_2| z^n), {\tilde X}_2^n \!\in\! {\cal T}_\epsilon^{(n)} (X_2| x_1^n, z^n) \!\right\}\!, \nonumber \\
& p_3 \!=\! P \left\{\! (X_1^n, X_2^n) \!\in\! {\cal T}_\epsilon^{(n)} (X_1, X_2| z^n), {\tilde X}_1^n \!\in\! {\cal T}_\epsilon^{(n)} (X_1| x_2^n, z^n) \!\right\}\!, \nonumber \\
& p_4 \!=\! P \left\{ (X_1^n, X_2^n), ({\tilde X}_1^n, {\tilde X}_2^n) \!\in\! {\cal T}_\epsilon^{(n)} (X_1, X_2| z^n) \right\} \!=\! p_1^2,
\end{align}
where ${\cal T}_\epsilon^{(n)} (X_2| x_1^n, z^n)$ and ${\cal T}_\epsilon^{(n)} (X_1| x_2^n, z^n)$ are defined in (\ref{T_conditional2}) and (\ref{T_conditional3}), respectively.
Since $X_1^n$, $X_2^n$ and ${\tilde X}_2^n$ are independent, we have
\begin{align}\label{p_2_up}
p_2 & = \sum_{(x_1^n, x_2^n) \in {\cal T}_\epsilon^{(n)} \!(X_1, X_2| z^n)} \!\!\!p(x_1^n) p(x_2^n) \sum_{ {\tilde x}_2^n \in {\cal T}_\epsilon^{(n)} \!(X_2| x_1^n, z^n)} \!\!\!p({\tilde x}_2^n)\nonumber\\
& \leq 2^{-n(I(X_1, X_2; Z) - 3 \epsilon)} 2^{n(H(X_2| X_1, Z) + \epsilon)} 2^{-n(H(X_2) - \epsilon)}\nonumber\\
& = 2^{-n(I(X_1, X_2; Z) + I(X_2; Z| X_1) - \delta_1 (\epsilon))}.
\end{align}
Similarly, $p_3$ can be upper bounded as follows
\begin{equation}\label{p_3_up}
p_3 \leq 2^{-n(I(X_1, X_2; Z) + I(X_1; Z| X_2) - \delta_1 (\epsilon))}.
\end{equation}

By introducing indicator variable
\begin{equation}\label{indicator_E0}
E' (l_1, l_2) \!=\! \left\{\!\!\!
\begin{array}{ll}
1,&  \!\!\!{\text {if}} ~(x_1^n (l_1), x_2^n (l_2)) \!\in\! {\cal T}_\epsilon^{(n)} (X_1, X_2| z^n),\\
0,&  \!\!\!{\text {otherwise}},\\
\end{array} \right.\!\!\!\!
\end{equation} 
where $z^n \in {\cal T}_\epsilon^{(n)} (Z)$ and $(l_1,l_2) \in {\cal L}_{1,m_1} \times {\cal L}_{2,m_2}$, $N\left(m_1, m_2, z^n\right)$ can be re-presented as
\begin{equation}\label{N0}
N(m_1, m_2, z^n) = \sum_{(l_1,l_2) \in {\cal L}_{1,m_1} \times {\cal L}_{2,m_2} } E' (l_1, l_2).
\end{equation}
Then, we have (\ref{N_expectation}) as follows and (\ref{N_square}) at the top of this page
\begin{align}\label{N_expectation}
{\mathbb E} \left\{N(m_1, m_2, z^n)\right\} = & \sum_{(l_1,l_2) \in {\cal L}_{1,m_1} \times {\cal L}_{2,m_2} } {\mathbb E} \left\{ E' (l_1, l_2) \right\}\nonumber\\
= & \sum_{(l_1,l_2) \in {\cal L}_{1,m_1} \times {\cal L}_{2,m_2} } p_1\nonumber\\
= & |{\cal L}_{1,m_1} \times {\cal L}_{2,m_2}| p_1\nonumber\\
= & 2^{n(R_1^{\text o} + R_1^{\text g} + R_2^{\text o} + R_2^{\text g})} p_1\nonumber\\
\leq & 2^{n (\Delta + \delta_1 (\epsilon))}.
\end{align}
According to (\ref{N_expectation}) and (\ref{N_square})
\setcounter{equation}{66}
\begin{align}\label{VarN1}
&{\text {Var}} \left\{ N(m_1, m_2, z^n) \right\} \nonumber\\
= & {\mathbb E} \left\{[N(m_1, m_2, z^n)]^2\right\} - \left\{{\mathbb E} [N(m_1, m_2, z^n)]\right\}^2\nonumber\\
\leq & 2^{n (\Delta + \delta_1 (\epsilon))} + \sum_{k \in \cal K} 2^{n (2 \Delta - \Delta_k + \delta_1 (\epsilon))}.
\end{align}
Theorem \ref{theorem3} is thus proven.

\bibliographystyle{IEEEtran}
\bibliography{IEEEabrv,Ref}

\begin{thebibliography}{1}
\providecommand{\url}[1]{#1}
\csname url@samestyle\endcsname
\providecommand{\newblock}{\relax}
\providecommand{\bibinfo}[2]{#2}
\providecommand{\BIBentrySTDinterwordspacing}{\spaceskip=0pt\relax}
\providecommand{\BIBentryALTinterwordstretchfactor}{4}
\providecommand{\BIBentryALTinterwordspacing}{\spaceskip=\fontdimen2\font plus
\BIBentryALTinterwordstretchfactor\fontdimen3\font minus
  \fontdimen4\font\relax}
\providecommand{\BIBforeignlanguage}[2]{{%
\expandafter\ifx\csname l@#1\endcsname\relax
\typeout{** WARNING: IEEEtran.bst: No hyphenation pattern has been}%
\typeout{** loaded for the language `#1'. Using the pattern for}%
\typeout{** the default language instead.}%
\else
\language=\csname l@#1\endcsname
\fi
#2}}
\providecommand{\BIBdecl}{\relax}
\BIBdecl

\bibitem{yang2015safeguarding}
N.~Yang, L.~Wang, G.~Geraci, M.~Elkashlan, J.~Yuan, and M.~Di~Renzo,
  ``Safeguarding {5G} wireless communication networks using physical layer
  security,'' \emph{IEEE Commun. Mag.}, vol.~53, no.~4, pp. 20--27, Apr. 2015.

\bibitem{wyner1975wire}
A.~D. Wyner, ``The wire-tap channel,'' \emph{Bell Sys. Tech. J.}, vol.~54,
  no.~8, pp. 1355--1387, Oct. 1975.

\bibitem{leung1978gaussian}
S.~Leung-Yan-Cheong and M.~Hellman, ``The gaussian wire-tap channel,''
  \emph{IEEE Trans. Inf. Theory}, vol.~24, no.~4, pp. 451--456, July 1978.

\bibitem{csiszar1978broadcast}
I.~Csisz{\'a}r and J.~K{\"o}rner, ``Broadcast channels with confidential
  messages,'' \emph{IEEE Trans. Inf. Theory}, vol.~24, no.~3, pp. 339--348, May
  1978.

\bibitem{tekin2008gaussian}
E.~Tekin and A.~Yener, ``The gaussian multiple access wire-tap channel,''
  \emph{IEEE Trans. Inf. Theory}, vol.~54, no.~12, pp. 5747--5755, Dec. 2008.

\bibitem{ekrem2008secrecy}
E.~Ekrem and S.~Ulukus, ``On the secrecy of multiple access wiretap channel,''
  in \emph{Proc. 46th Allerton Conf. Commun., Contr., Comput.}, Illinois, USA,
  Sep. 2008, pp. 1014--1021.

\bibitem{tekin2008general}
E.~Tekin and A.~Yener, ``The general gaussian multiple-access and two-way
  wiretap channels: Achievable rates and cooperative jamming,'' \emph{IEEE
  Trans. Inf. Theory}, vol.~54, no.~6, pp. 2735--2751, June 2008.

\bibitem{nafea2019generalizing}
M.~Nafea and A.~Yener, ``Generalizing multiple access wiretap and wiretap {II}
  channel models: Achievable rates and cost of strong secrecy,'' \emph{IEEE
  Trans. Inf. Theory}, vol.~65, no.~8, pp. 5125--5143, Aug. 2019.

\bibitem{el2011network}
A.~El~Gamal and Y.-H. Kim, \emph{Network information theory}.\hskip 1em plus
  0.5em minus 0.4em\relax {Cambridge University Press}, 2011.

\end{thebibliography}

\end{document}